\documentclass[preprint,showpacs,preprintnumbers,amsmath,amssymb]{revtex4}

\usepackage{graphicx}

\begin{document}

\title{On the Relativistic Invariance of Entanglement}

\author{Esteban Castro-Ruiz$^{1,2}$, Eduardo Nahmad-Achar$^{1}$}

\affiliation{$^{1}$Instituto de Ciencias Nucleares, Universidad Nacional Aut\'onoma de M\'exico,
Apdo. Postal 70-543 M\'exico 04510 D.F. \\ $^{2}$Facultad de Ciencias, Universidad Nacional Aut\'onoma de M\'exico, Apdo. Postal 70-542 M\'exico 04510 D.F.}

\

\begin{abstract}

In this work we study  the entanglement properties under a Lorentz boost of a pair of spin-1 massive particles, with spin and momentum as the sole degrees of freedom of the system. Different cases for entanglement between spins and momenta are considered as initial states and it is shown that entanglement  between the different degrees of freedom is, in general, not invariant.  We analyze the entanglement change in terms of the Wigner rotation induced by the transformation and show that, for the case of spin-1 particles, the relative entanglement change between different states depends strongly on the boost velocity.

\


\end{abstract}


\maketitle

\section{Introduction}

Measurement on one part of an entangled system changes the probability of the outcome of measurements on the second part of the system, according to quantum mechanics. This assumes a time-ordering of the events, a concept which is not frame-independent for spatially separated events, according to special relativity. In order to understand this in terms of a concrete example, it is convenient to study entanglement from a relativistic point of view. In particular, we address the question of entanglement invariance under Lorentz transformations and the related invariance of quantum non-local effects. We study how the entanglement change depends on the spin of the particles. This is done in relation to previous research carried out for the spin-1/2 case, in which Friis et al.~\cite{Friis} found that entanglement is not conserved with respect to some particular decompositions of the Hilbert space. On the other hand, Alsing et al.~\cite{Alsing} showed that each single-particle subsystem, {\it A} and {\it B}, undergoes a local unitary transformation under a Lorentz boost; hence conserving entanglement with respect to the $A$ vs. $B$ partition. 

Here we calculate the entanglement of a two-particle spin-1 system before and after a Lorentz boost, considering the momentum degree of freedom as a discrete two-level variable $p_{+}$  and $p_{-}$ (i.e., only direction of movement matters). Therefore, only sharp momentum distributions are studied; for an analysis of a more realistic wave-packet distributions see~\cite{Gingrich,Kamil}.

\section{Description of the Problem}

We analyze a system of two spin-one particles propagating along the $z$ axis, and compare the entanglement before and after a Lorentz boost on the state. The boost is taken to be along the $x$ axis, as only components of boosts perpendicular to the propagation direction generate a change in entanglement~\cite{Alsing}.

Since momentum states are considered as two-level systems, we work in a 
$2\times 3\ \times\ 2\times 3$ dimensional Hilbert space
\begin{eqnarray}
H &=& H_{p}\otimes H_{s} \nonumber \\
&=&(H_{A}\otimes H_{B})_{p}\otimes(H_{A}\otimes H_{B})_{s}
\end{eqnarray}

There are four physically distinct subspaces, each one related to the spin ($s$) or momentum ($p$) degrees of freedom for observers $A$ and $B$: $s_A$, $p_A$, $s_B$, $p_B$. We calculate the change of entanglement for different decompositions of the total Hilbert space, as done by~\cite{Friis}. The reduced Hilbert spaces are obtained from the total space by tracing over the relevant degrees of freedom.

First we will consider the $s$ vs. $p$ partition, obtained by tracing over the spin or the momentum degrees of freedom, and will show that entanglement is not invariant under Lorentz boosts in this case. 
Next, we trace over all subspaces except one to obtain a partition of $1$ vs. $3$ subsystems. Here we calculate the total change of entanglement by adding up the contributions of all possible decompositions of this type. 
The two remaining partitions are called the $A$ vs. $B$, and the mixed partitions. They are obtained by tracing over the $A$ (or $B$) subspaces, and over the $s_A$ and $p_B$ (or the $p_A$ and $s_B$) subspaces, respectively. Entanglement is conserved in these two last cases.

We choose the initial state to be separable with respect to the $s$ vs. $p$ partition{\label{bbb}}
\begin{equation}
 \vert\psi\rangle = \vert s\rangle \otimes \vert p\rangle,
 \end{equation}
where we parametrize the momentum state as 
 \begin{equation}{\label{momentumstate}}
 \vert p\rangle = \cos\alpha\, \vert p_+,p_-\rangle + \sin\alpha\,\vert p_-,p_+\rangle 
 \end{equation}
Note that this momentum state parametrization only makes sense for distinguishable particles, as two mass-distiguishable spin-$1$ atoms. (If one were to take $\vert p\rangle = (\cos\alpha\, \vert p_+,p_-\rangle + e^{i\phi} \sin\alpha\,\vert p_-,p_+\rangle)$, the partial traces of the density matrix $\rho = \vert\psi\rangle\langle\psi\vert$ in {\it any} partition taken below will remain unaltered with respect to this parametrization; thus, only one parameter $\alpha$ suffices.)

For spin states, we consider two different parametrizations: 
\begin{eqnarray}
\vert s_1\rangle = \sin\theta\cos\phi\,\vert 1 1\rangle+\sin\theta\sin\phi\,\vert 00\rangle+\cos\theta\,\vert -1-1\rangle{\label{1s}}\\
\vert s_2\rangle = \sin\theta\cos\phi\,\vert 1 -1\rangle+\sin\theta\sin\phi\,\vert -11\rangle+\cos\theta\,\vert 00\rangle{\label{2s}}
\end{eqnarray}
where the first element in the {\it ket} relates to the first particle's spin projection, and the second element in the {\it ket} relates to that of the second particle.  For a further $3$-dimensional parametrization see~\cite{Castro}.

Entanglement change is then calculated for whole classes of initial states, depending on the values of the parameters. 
Since all states considered are pure, the linear entropy
\begin{equation}
E = \sum_{i}\ (1 - Tr\left(\rho_{i}^2\right)\,)
\end{equation}
is used as an entanglement measure.

The relativistic transformation of spin and momentum states is given by
\begin{equation} 
U(\Lambda)\,\vert p,\,s\rangle=\sum_{s^{\prime}}\,D^{(j)}(W(\Lambda,\,p))\,\vert\Lambda p,\,s^{\prime}\rangle
\end{equation}
The coefficients $D^{(j)}(W(\Lambda,\,p))$ are the spin-$j$ representation of the rotation group $SO(3)$~\cite{Weinberg}. In the case of spin-$1$ particles, with momentum along the $z$ and $-z$ directions, and a boost $\Lambda$ in the direction of the $x$ axis, the rotation matrices are given in terms of the Wigner angle $\Omega$, which is a function of the standard boost rapidity $\xi$, and the $\Lambda$ boost rapidity $\eta$ 
\begin{equation} 
\tanh\Omega=\frac{\sinh\xi\,\sinh\eta}{\cosh\xi+\cosh\eta}
\end{equation} 
The rotation of the spin state $W(\Lambda,\,\pm p)$ is momentum-dependent: it is a rotation by an angle $\Omega$ along the $y$ axis, for $p_{+}$ and a rotation by the same amount but in the opposite direction, for momentum  $p_{-}$. The momentum states for the different particles thus suffer different transformations and the whole state $\vert\psi\rangle$ becomes entangled with respect to the $s$ vs. $p$ partition. A complete derivation can be found in~\cite{Castro}.

\section{Results}

In agreement with~\cite{Friis,Alsing}, we found that entanglement with respect to the $A$ vs. $B$ and with respect to the mixed partition is conserved for all states. On the other hand, entanglement with respect to the $s$ vs. $p$ partition and the $1$ vs. $3$ partition is not conserved in general. The dependence of the entanglement change, $\Delta E$, on the initial entanglement of the momentum state $\vert p\rangle$ (eqn.(\ref{momentumstate})) is easy to analyze:  $\Delta E$ is zero for all separable momentum states in which $\alpha= \frac{n\pi}{2} $ and is maximized for maximally entangled momentum states. The function $\Delta E = \Delta E(\theta,\phi)$ changes only by a scale factor when the parameter $\alpha$ varies.

\begin{figure}[h]
\scalebox{0.750}{\includegraphics{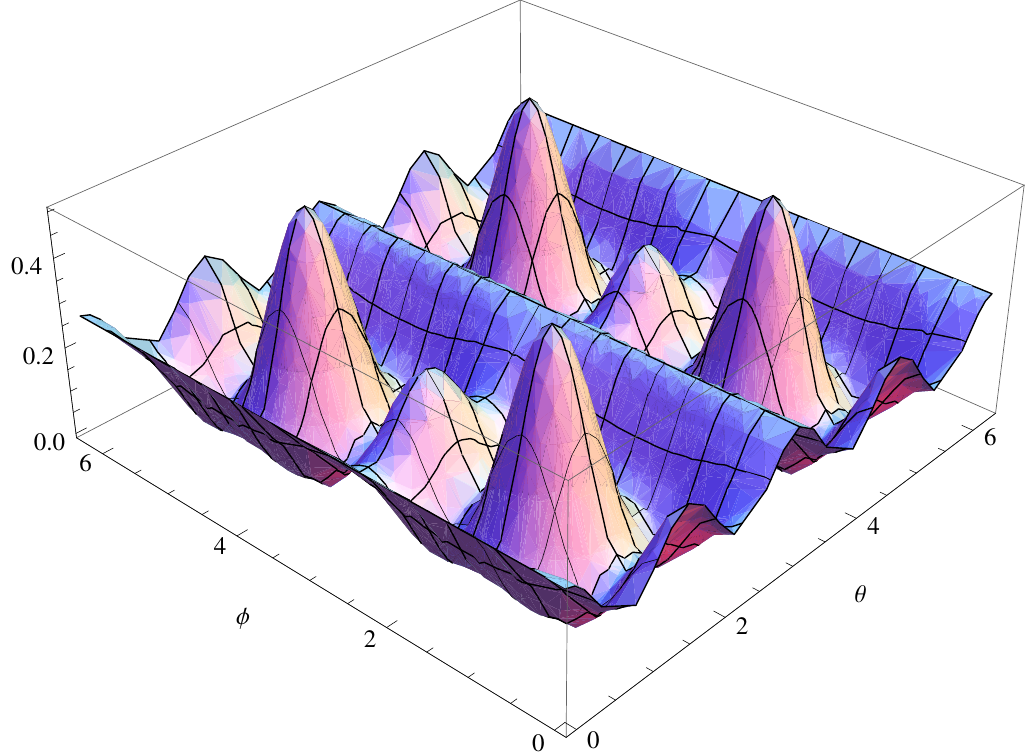}}

\

\scalebox{0.750}{\includegraphics{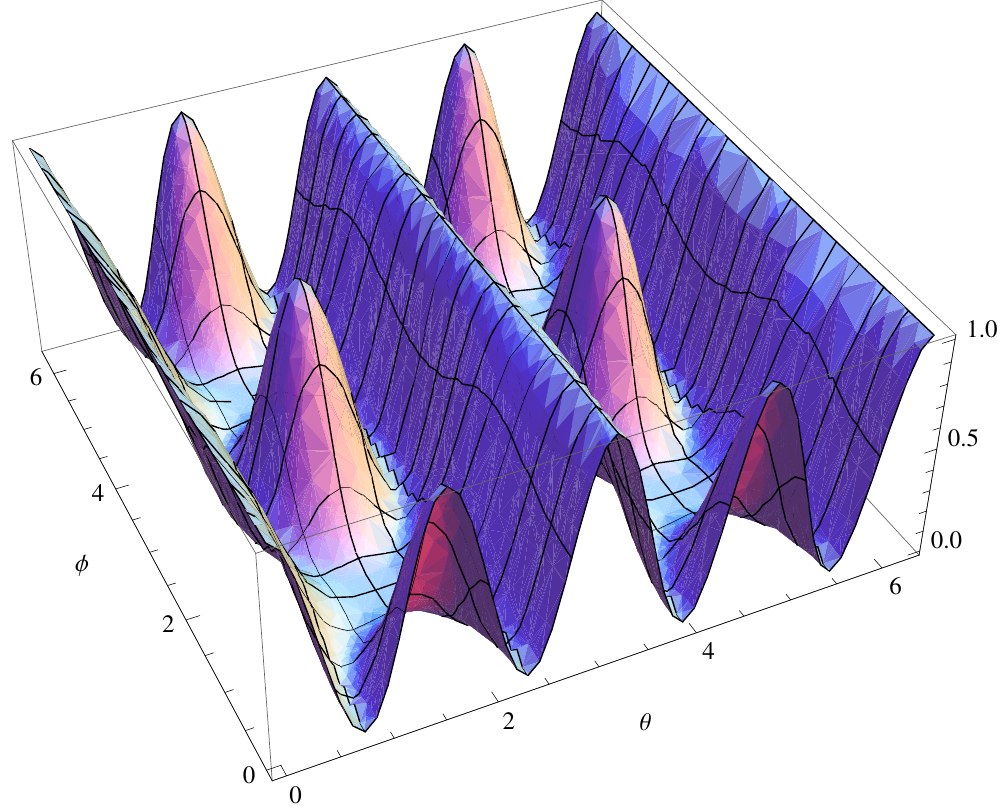}}
\caption{$\Delta E$ as a function of $\theta$ and $\phi$ for a Wigner angle of $\frac{\pi}{8}$ (above) and $\frac{\pi}{2}$ (below), in the $1$ vs. $3$ partition. Maxima are found for states $\vert 1 1\rangle\pm\vert -1-1\rangle$ and $\vert 00\rangle$ respectively.}
\label{figura1}
\end{figure}

On the other hand, the dependence of $\Delta E$ on the Wigner angle, $\Omega$, is complicated. For example, consider the state $\vert s_{1}\rangle$ of eqn.(\ref{1s}) in the $1$ vs. $3$ partition. For a relatively small Wigner angle ($\Omega = \frac{\pi}{8}$), the states that exhibit the maximal entanglement change are $\vert 1 1\rangle\pm\vert -1-1\rangle$. For $\Omega = \frac{\pi}{2}$,  corresponding to the limit of the speed of light, $\Delta E = 0$ for the above mentioned states and is maximal for the state  $\vert 00\rangle$, as seen in fig(\ref{figura1}). The minima of the entanglement change, for this parametrization and partition correspond to the maximally entangled states $\frac{1}{\sqrt{3}}[\vert 1 1\rangle\pm\vert 00\rangle\,\pm\vert -1-1\rangle]$. 

There seems to be a relation between the initial spin entanglement and the change of entanglement with respect to the $1$ vs. $3$ partition: separable sates undergo the maximal change in entanglement while maximally entangled states don't change their entanglement at all. However, this cannot be considered as a general rule since $\Delta E$ depends significantly on the Wigner angle for the separable states.

\begin{figure}[h]
\scalebox{0.75}{\includegraphics{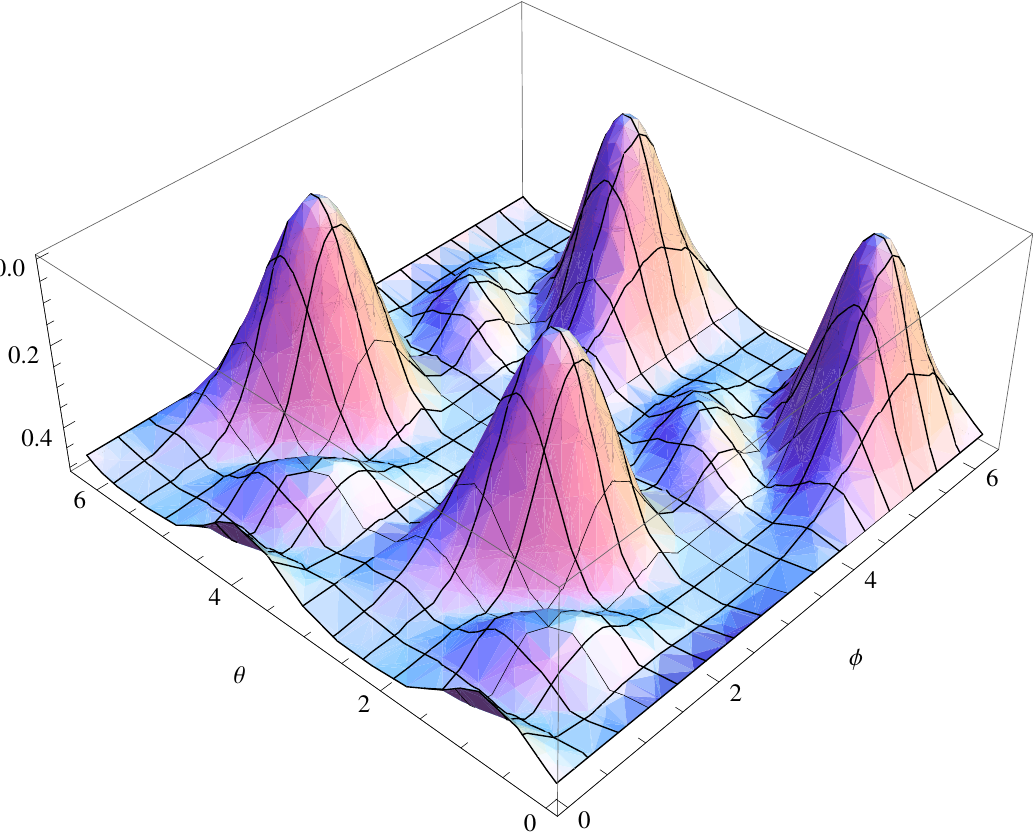}}
\caption{$\Delta E$ as a function of $\theta$ and $\phi$ for a Wigner angle of $\frac{\pi}{4}$, in the $1$ vs. $3$ partition. The plot is shown upside down, to look at the minima. These minima are stable against variations in $\Omega$.}
\label{figura3}
\end{figure}

Consider next the same parametrization $\vert s_1\rangle$ eqn.(\ref{1s}) now with the partition $s$ vs. $p$. As shown in fig(\ref{figura3}), the only state whose entanglement remains invariant is 
$\frac{1}{\sqrt{3}}[\vert 1 1\rangle\pm\vert 00\rangle\,\pm\vert -1-1\rangle]$. In fact, this state remains invariant under all transformations of the form  $D(W(\Lambda, p))\otimes D(W(\Lambda, -p))$, a fact that makes it interesting for quantum communication processes in a relativistic framework.

\begin{figure}[h]
\scalebox{0.75}{\includegraphics{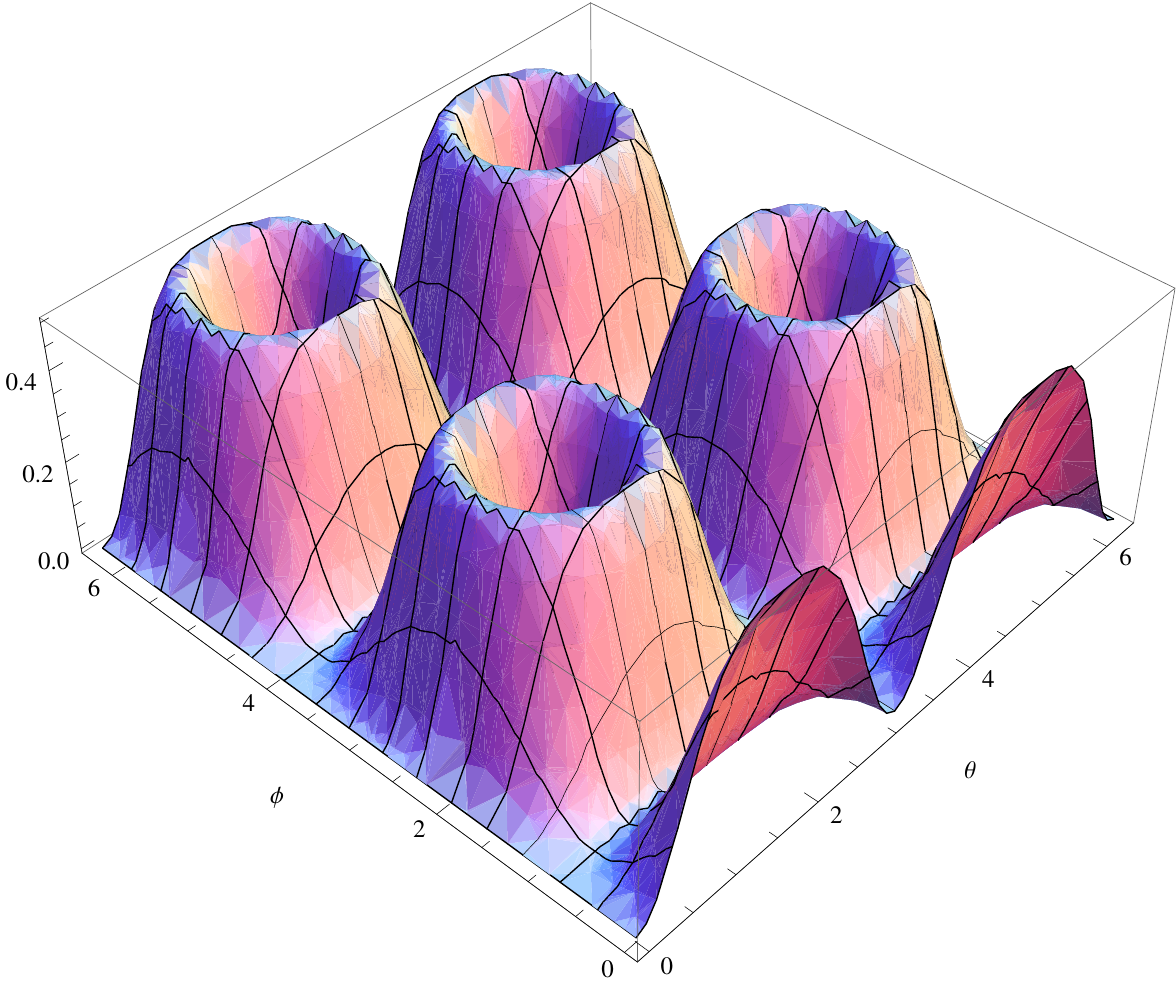}}
\caption{$\Delta E$ as a function of $\theta$ and $\phi$ for a Wigner angle of $\frac{\pi}{2}$, in the second spin parametrization and the $s$ vs. $p$ partition. The minima correspond to the Bell-type state $\frac{1}{\sqrt{2}}[\,\vert 1\,-1\rangle-\vert -1\,1\rangle]$.}
\label{figura4}
\end{figure}

If we analyze Bell-Type states of just two spin polarizations of our spin-1 particles, we see that it is possible to increase their entanglement by means of a Lorentz Boost. While this might seem obvious from the fact that we have a larger dimensional Hilbert space, it is worth noting that this is also strongly dependent on the magnitude of the boost, as we can see from fig(\ref{figura4}), where the state $\frac{1}{\sqrt{2}}[\,\vert 1\,-1\rangle-\vert -1\,1\rangle]$ behaves notoriously as a maximally entangled state for $\Omega=\frac{\pi}{2}$, in which case it has $\Delta E = 0$.

\section{Conclusion}

Entanglement is conserved for the $A$ vs. $B$ and the mixed partitions, due to the unitarity of the single-particle transformations. For the remaining partitions, entanglement is not invariant for most of the states. The behavior of the entanglement change depends on the initial momentum entanglement through a scale factor on the whole set of states. The dependence on the Wigner angle is much more interesting since, for instance, it is possible to change the set of maxima and minima of the $\Delta E$ function by means of an increment on the boost velocity. However, the partitions for which this statement is valid offer an entanglement which is of no use for quantum communication, since it is not associated with any nonlocal effect. 

It is worth stressing that the current definition of entanglement, and the measure of it, is not frame-independent. Moreover, its dependence on the inertial observer is complex and rich.

\acknowledgments

This work was partially supported by FONCICYT (project 94142) and DGAPA-UNAM.

\end{document}